\documentclass[twocolumn]{fairmeta}

\usepackage{amsmath}
\usepackage{amssymb}
\usepackage{enumitem}
\usepackage{subcaption}
\usepackage{tikz}
\usetikzlibrary{positioning, arrows.meta, shadows, backgrounds, calc, shapes.geometric, fit}
\usepackage{balance}

\title{RankGraph-2: Lifecycle Co-Design for Billion-Node Graph Learning in Recommendation}

\author[1]{Renzhi Wu}
\author[1]{Zikun Cui}
\author[1]{Junjie Yang}
\author[1]{Tai Guo}
\author[1]{Hong Li}
\author[2]{Xian Chen}
\author[3]{Li Yu}
\author[3]{Ke Pan}
\author[1]{Sri Reddy}
\author[1]{Mahesh Srinivasan}
\author[1]{Nipun Mathur}
\author[3]{Haomin Yu}
\author[1]{Hong Yan}

\affiliation[1]{Meta MRS}
\affiliation[2]{Meta Training Data Infra}
\affiliation[3]{FB Monetization}
\abstract{Graph-based retrieval at billion-node scale requires jointly solving three tightly coupled problems---graph construction, representation learning, and real-time serving---yet existing work addresses each in isolation. We present RankGraph-2, a framework deployed at Meta that co-designs all three lifecycle stages for similarity-based retrieval (U2U2I and U2I2I), where each stage's requirements shape the others. Serving requires a co-learned cluster index to avoid expensive online KNN---this pushes index co-training into the training objective. Training benefits from the observation that similarity-based retrieval tolerates pre-computed neighborhoods, eliminating online graph infrastructure---this requires construction to produce self-contained data. Construction must also support hour-level refresh for item coverage. Acting on these cascading requirements, RankGraph-2 reduces hundreds of trillions of edges to hundreds of billions via subsampling with popularity bias correction, pre-computes multi-hop neighborhoods via personalized PageRank, and co-learns a residual-quantization cluster index that reduces serving computational cost by 83\%. This lifecycle co-design enables a simple architecture to achieve 3.8$\times$ higher recall than a GAT + Deep Graph Infomax model on a bipartite graph and 2.1$\times$ higher than PyTorch-BigGraph on item retrieval.
RankGraph-2 delivers up to +0.96\% CTR and +2.75\% CVR, and has powered \textbf{20+ retrieval launches} across major surfaces.}

\date{\today}
\correspondence{Renzhi Wu at \email{renzhiwu@meta.com}}

\metadata[Keywords]{Recommendation system, graph learning, graph neural network}

\begin{document}

\maketitle

\section{Introduction}
\label{sec:intro}
Graph-based retrieval---using learned embeddings on user-item graphs for candidate generation---is a critical component of modern recommendation systems. Deploying graph neural networks (GNNs) for this purpose at billion-node scale requires solving three tightly coupled problems: constructing the graph, learning representations on it, and serving those representations in real time. These three stages form a \emph{lifecycle}, and a bottleneck in any one stage limits the entire system. Yet existing work overwhelmingly focuses on a single stage in isolation. Academic GNN research optimizes model architectures on small, static graphs~\citep{zhang2019heterogeneous, chen2023heterogeneous, he2020lightgcn}. Industrial systems such as GiGL~\citep{zhao2025gigl}, LiGNN~\citep{borisyuk2024lignn}, and GraphScale~\citep{gupta2024graphscale} advance training infrastructure but assume the input graph is given and do not address serving cost. Neither community has systematically addressed the question: \emph{how should graph construction, training, and serving be co-designed when the graph has billions of nodes and hundreds of trillions of potential edges?}

This gap matters for retrieval in practice. At Meta, naively constructing the full co-engagement graph is infeasible; naively training on it requires expensive online graph infrastructure; and naively serving the resulting embeddings via online KNN requires thousands of machines. Each stage imposes requirements on the others: graph construction determines what signals training can learn from, the training objective determines what the embeddings capture, and serving cost determines whether the system can be deployed at all. Optimizing any one stage without considering the others leads to suboptimal outcomes.

\begin{figure}[htbp!]
    \centering
    \resizebox{0.98\columnwidth}{!}{%
    \begin{tikzpicture}[
        >=Stealth,
        font=\sffamily,
        stage/.style={
            rectangle, rounded corners=8pt, thick,
            minimum width=3.0cm, minimum height=1.2cm,
            align=center, font=\normalsize\bfseries,
            drop shadow={opacity=0.1, shadow xshift=1.5pt, shadow yshift=-1.5pt}
        },
        rlabel/.style={
            font=\small, align=right, text width=2.0cm,
            text=red!75!black, fill=white, fill opacity=0.9, text opacity=1,
            inner sep=3pt, rounded corners=2pt
        },
        slabel/.style={
            font=\small, align=left, text width=2.0cm,
            text=blue!75!black, fill=white, fill opacity=0.9, text opacity=1,
            inner sep=3pt, rounded corners=2pt
        },
        reqarr/.style={
            ->, thick, red!70!black, dashed, rounded corners=8pt
        },
        solarr/.style={
            ->, thick, blue!70!black, rounded corners=8pt
        },
        mainflow/.style={
            ->, line width=1.5pt, gray!60
        }
    ]

    \node[stage, fill=blue!6, draw=blue!50!black] (G) at (0, 0) {Graph\\Construction};
    \node[stage, fill=teal!6, draw=teal!50!black] (T) at (0, -3.2) {Model\\Training};
    \node[stage, fill=orange!6, draw=orange!50!black] (S) at (0, -6.4) {Real-time\\Serving};

    \draw[mainflow] (G.south) -- node[fill=white, inner sep=3pt, font=\small\sffamily, text=gray!80!black, align=center] {edge-centric \\ data} (T.north);
    \draw[mainflow] (T.south) -- node[fill=white, inner sep=3pt, font=\small\sffamily, text=gray!80!black, align=center] {embeddings \\ + index} (S.north);

    \def\rxin{-1.95}   
    \def\rxout{-4.25}  
    \def\lxin{1.95}    
    \def\lxout{4.25}   
    \def\ys{0.15cm}   


    \draw[reqarr] ([yshift=\ys]S.west) -- (\rxin, -6.25) |- ([yshift=-\ys]T.west);
    \node[rlabel, left=2pt] at (\rxin, -4.8) {Low cost:\\cluster-index\\to replace\\online KNN};

    \draw[reqarr] ([yshift=\ys]T.west) -- (\rxin, -3.05) |- ([yshift=-\ys]G.west);
    \node[rlabel, left=2pt] at (\rxin, -1.6) {Low cost:\\removing\\graph infra};

    \draw[reqarr] ([yshift=-\ys]S.west) -- (\rxout, -6.55) |- ([yshift=\ys]G.west);
    \node[rlabel, left=2pt] at (\rxout, -3.2) {High fresh\\item coverage};


    \draw[solarr] ([yshift=-\ys]G.east) -- (\lxin, -0.15) |- ([yshift=\ys]T.east);
    \node[slabel, right=2pt] at (\lxin, -1.6) {Pre-computed\\neighbors in\\self-contained\\format};

    \draw[solarr] ([yshift=-\ys]T.east) -- (\lxin, -3.35) |- ([yshift=\ys]S.east);
    \node[slabel, right=2pt] at (\lxin, -4.8) {Co-learned\\cluster index};

    \draw[solarr] ([yshift=\ys]G.east) -- (\lxout, 0.15) |- ([yshift=-\ys]S.east);
    \node[slabel, right=2pt] at (\lxout, -3.2) {Graph\\rebuild\\in $<$1 hour};

    \node[font=\footnotesize\sffamily, draw=gray!30, fill=gray!4, rounded corners=4pt, inner sep=4pt] at (0, -8.2) {
        \begin{tabular}{l}
        \tikz[baseline=-0.5ex]{\draw[reqarr] (0,0) -- (0.5,0);} Requirement (Upstream) \\
        \tikz[baseline=-0.5ex]{\draw[solarr] (0,0) -- (0.5,0);} Co-designed Solution (Downstream)
        \end{tabular}
    };

    \end{tikzpicture}
    }
    \caption{Lifecycle co-design in RankGraph-2. Serving and training impose requirements (\textcolor{red!70!black}{dashed red}, left) on upstream stages. Each requirement is resolved by a co-designed solution (\textcolor{blue!70!black}{solid blue}, right), creating mutual dependencies across the pipeline.}
    \label{fig:lifecycle}
\end{figure}
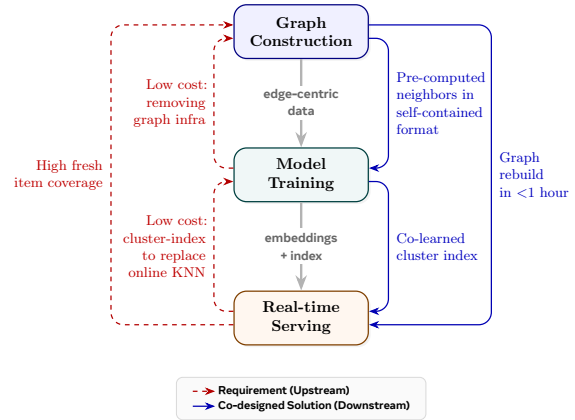

We present RankGraph-2, a unified framework built on the principle that \textbf{lifecycle co-design---letting each stage's requirements shape the others---unlocks performance that optimizing stages independently cannot achieve}. RankGraph-2 builds upon our original RankGraph system~\citep{wu2025rankgraphunifiedheterogeneousgraph} and achieves 3.8$\times$ higher recall than a GAT + Deep Graph Infomax model~\citep{velickovic2018graph, velickovic2019deep} on a bipartite graph, and 2.1$\times$ higher than PyTorch-BigGraph~\citep{lerer2019pytorch} on item retrieval, with a simpler architecture (Section~\ref{sec:experiments}). We focus on similarity-based retrieval---U2U2I (user-to-user-to-item) and U2I2I (user-to-item-to-item)---where we empirically find graph-based embeddings are most effective, since these retrievals rely on structural similarity rather than the real-time behavioral features that dominate direct user-to-item matching. Below, we trace how each lifecycle stage creates requirements that cascade to the others, and how co-design resolves them (Figure~\ref{fig:lifecycle}):

\textit{(1) Serving: from online KNN to cluster-based retrieval.} For U2U2I, we empirically find that constraining the candidate pool to recently active users (e.g., past 15 minutes) yields the best quality. Online KNN over this constantly changing pool is prohibitively expensive at our traffic scale. Pre-computed static KNN mappings are also insufficient, since the active user set changes every few minutes. We instead adopt a cluster-based approach: each user is assigned a cluster, and each cluster maintains a queue of items from its recently active members. At serving time, U2U2I reduces to reading the latest items from the target user's cluster queue---far cheaper than KNN. However, a na\"ively learned index degrades retrieval quality; the index must be co-optimized with the embedding objective to match KNN accuracy. This \emph{requires training to co-learn the index}. Additionally, while the structural similarity signals that drive U2U and I2I retrieval are relatively stable, fresh item coverage remains important---\emph{requiring graph construction to support hour-level refresh}.

\textit{(2) Training: co-learning for serving, graph-infra-free by design.} Training addresses two cascading requirements. First, it co-learns a residual-quantization cluster index alongside the embeddings, so the model directly optimizes for serving quality rather than relying on post-hoc quantization. Second, we observe that for similarity-based retrieval, the structural signals that drive U2U and I2I similarity are less time-sensitive than those used in CTR prediction, enabling us to move neighborhood computation offline. This not only eliminates online graph infrastructure---no graph storage, no distributed sampling engines~\citep{gupta2024graphscale, zheng2022bytegnn, zhao2025gigl}---but actually enables \emph{higher-quality} neighborhoods: offline, we can run multi-hop PPR over the full graph and apply popularity bias correction, operations that would be infeasible under online latency constraints. This in turn \emph{requires graph construction to provide high-quality pre-computed neighbors in a self-contained format}.

\textit{(3) Graph construction: satisfying both downstream stages.} Construction receives requirements from both training (high-quality self-contained edge-centric data with pre-computed neighborhoods) and serving (hour-level refresh for item coverage). At Meta, the raw co-engagement graph has hundreds of trillions of edges. We develop an edge subsampling strategy with popularity bias correction that reduces the graph by three orders of magnitude while preserving retrieval-relevant structure. The pipeline pre-computes multi-hop neighborhoods via personalized PageRank (PPR) and packages everything into an edge-centric format ready for direct ingestion (Section~\ref{sec:graph_construction}). The entire pipeline completes within one hour, enabling the 3-hour refresh cycle that serving requires.

Deployed across multiple surfaces at Meta, RankGraph-2 achieves up to 3.8$\times$ higher offline recall than existing baselines, delivers up to +0.96\% CTR and +2.75\% CVR improvements in 14-day A/B tests, and reduces U2U serving infrastructure cost by 83\%.

Our contributions are:
\begin{itemize}[leftmargin=*]
    \item We identify lifecycle co-design---where serving, training, and construction requirements cascade across stages---as a key missing piece in billion-scale GNN systems for retrieval, and present RankGraph-2 as a concrete realization.
    \item We show that similarity-based retrieval (U2U2I, U2I2I) does not require online graph infrastructure: pre-computed neighborhoods yield quality comparable to online sampling, enabling GNN training on standard ML infrastructure without dedicated graph storage or sampling engines.
    \item We propose a heterogeneous co-engagement graph with all three edge types (U-U, I-I, U-I) derived from engagement data alone, with popularity bias correction for I-I edges, and PPR-based neighbor pre-computation that produces self-contained edge-centric training data.
    \item We develop a co-learned cluster index with novel regularization and biased code selection that prevents codebook collapse under continuous training, reducing serving cost by 83\%.
    \item We demonstrate that this co-design enables a simple architecture to achieve 3.8$\times$ higher recall than GAT + Deep Graph Infomax, and 2.1$\times$ higher than PyTorch-BigGraph on item retrieval, with +0.96\% CTR and +2.75\% CVR in A/B tests.
\end{itemize}
\section{Related Work}
\label{sec:related}

\begin{table*}[t]
\centering
\caption{Cross-stage design coupling in billion-scale recommendation systems. For competing systems, ``Cross-stage gaps'' highlights dependencies left unaddressed. For RankGraph-2, the same column shows the co-design couplings that close these gaps.}
\label{tab:comparison}
\resizebox{\textwidth}{!}{
\begin{tabular}{@{}lll@{}}
\toprule
\textbf{System} & \textbf{Stages optimized} & \textbf{Cross-stage gaps / couplings} \\
\midrule
GiGL~\citep{zhao2025gigl} & Training (in-memory sampling) & \begin{tabular}[t]{@{}l@{}}Graph assumed given; requires online graph\\infra; serving via ANN\end{tabular} \\[6pt]
LiGNN~\citep{borisyuk2024lignn} & Training (adaptive multi-hop) & \begin{tabular}[t]{@{}l@{}}Graph assumed given; requires online graph\\infra; serving via ANN\end{tabular} \\[6pt]
MacGNN~\citep{liu2024macgnn} & \begin{tabular}[t]{@{}l@{}}Training (macro-node\\aggregation)\end{tabular} & \begin{tabular}[t]{@{}l@{}}Graph construction loses micro-node\\precision; serving via ANN\end{tabular} \\[6pt]
\midrule
\textbf{RankGraph-2} & \begin{tabular}[t]{@{}l@{}}\textbf{All three stages,}\\
\textbf{jointly co-designed}\end{tabular} & \begin{tabular}[t]{@{}l@{}}Construction $\rightarrow$ Training: pre-computed\\PPR neighbors, edge-centric format\\Training $\rightarrow$ Serving: co-learned index\\Serving $\rightarrow$ Construction: 3h refresh cycle\end{tabular} \\
\bottomrule
\end{tabular}
}
\end{table*}

\noindent\textbf{GNNs for recommendation.}
GNN-based recommendation has progressed from homogeneous models (GCN~\citep{bruna2013spectral}, GraphSAGE~\citep{hamilton2017inductive}, LightGCN~\citep{he2020lightgcn}) to heterogeneous architectures (HetGNN~\citep{zhang2019heterogeneous}, HAN~\citep{wang2019heterogeneous}, HGT~\citep{hu2020heterogeneous}) that aggregate across node and edge types. These models are evaluated on small static graphs and focus on aggregation mechanisms. RankGraph-2 uses a simple heterogeneous aggregator but invests in lifecycle co-design---eliminating graph infrastructure and pushing serving constraints upstream---which we find to be a stronger lever than model complexity at scale. We compare against a GAT + Deep Graph Infomax baseline~\citep{velickovic2018graph, velickovic2019deep} and PyTorch-BigGraph~\citep{lerer2019pytorch} in Section~\ref{sec:experiments}.

\noindent\textbf{Graph construction for recommendation.}
Most work constructs a bipartite user-item graph from interactions~\citep{zheng2018spectral, sharma2024survey, wu2022graph}. Some enrich it with social edges~\citep{salamat2021heterographrec, yang2022large} or knowledge graph relations~\citep{tien2020graph, xia2023disentangled}. RankGraph-2 instead derives all three edge types (U-U, I-I, U-I) from engagement data alone, without requiring external data sources. This ensures direct alignment with recommendation-relevant signals and enables popularity bias correction on I-I edges---a technique we find critical for preventing popular items from dominating the learned representations (Section~\ref{sec:ablation}).

\noindent\textbf{Scalable training and billion-scale systems.}
Neighborhood sampling methods (PinSage~\citep{ying2018graph}, NIA-GCN~\citep{sun2020neighbor}, ByteGNN~\citep{zheng2022bytegnn}) and distributed frameworks (GraphScale~\citep{gupta2024graphscale}, DGL~\citep{wang2019deep}, PyG~\citep{fey2025pyg}, AliGraph~\citep{zhu2019aligraph}) focus on making training efficient, but assume the graph is given.
Recent billion-scale systems address deployment but make different architectural trade-offs.
GiGL~\citep{zhao2025gigl} advocates for real-time in-memory subgraph sampling, requiring substantial infrastructure for graph storage and distributed sampling engines. We find that for similarity-based retrieval, this infrastructure is unnecessary: pre-computed neighborhoods yield comparable quality, allowing RankGraph-2 to train on standard ML infrastructure without dedicated graph systems.
LiGNN~\citep{borisyuk2024lignn} effectively models temporal heterogeneous graphs at LinkedIn but relies entirely on approximate nearest neighbor search for serving, which incurs substantial infrastructural overhead. RankGraph-2 introduces a co-learned cluster index to eliminate online KNN entirely.
MacGNN~\citep{liu2024macgnn} tackles neighbor explosion by clustering micro-nodes into macro-nodes. While computationally elegant, this inherently sacrifices micro-level relational precision---tail users and niche items are forced into generalized macro-clusters. RankGraph-2's edge subsampling protocol preserves the exact topological location of every individual node while achieving tractability through selective edge compression with popularity bias correction.
Other approaches take fundamentally different architectures: OneRec~\citep{wang2024onerec} replaces the retrieve-and-rank cascade with a unified generative model, and PinFM~\citep{zhou2024pinfm} pretrains a foundation model on user activity sequences. While these sequential approaches excel at capturing short-term temporal intent, they struggle to model explicit multi-hop collaborative filtering structures. We view RankGraph-2 not as a competitor to generative models, but as a complementary structural prior for downstream sequences.

\noindent\textbf{Embedding-based retrieval serving.}
Approximate KNN with index structures such as HNSW~\citep{malkov2018efficient}, Faiss~\citep{douze2024faiss}, and Milvus~\citep{wang2021milvus} is the standard serving approach. Recent work combines KNN with clustering for better control~\citep{zhang2023divide}. However, approximate KNN still requires substantial infrastructure at high traffic volumes. RankGraph-2 replaces online KNN with a co-learned cluster index, reducing serving cost by 83\%.

\noindent\textbf{Positioning.}
Existing systems optimize individual stages but leave cross-stage dependencies unaddressed: training-focused systems require online graph infrastructure and serve via ANN; generative models bypass graph structure entirely. RankGraph-2 co-designs all three stages so that each stage's output is tailored to the others' requirements, with serving, training, and construction requirements cascading across the pipeline. Table~\ref{tab:comparison} contrasts the cross-stage coupling of each system.

\section{Problem Setup}
Given user-item interactions data $D$ = \{($\text{user}_0$, $\text{item}_0$, $\text{interaction}_0$), ($\text{user}_1$, $\text{item}_1$, $\text{interaction}_1$), ...\}, our objectives are:
\begin{enumerate}[leftmargin=*]
    \item \textbf{To generate high-quality embeddings} for all users and items present in the interaction data $D$;
    \item \textbf{To efficiently serve these embeddings} for the retrieval stage in recommendation systems, supporting scenarios such as user-to-user-to-item (U2U2I) and user-to-item-to-item (U2I2I) retrieval.
\end{enumerate}

We focus on graph-based approaches: a graph construction module $g$ creates $G = g(D)$, and a graph learning model $M$ generates embeddings. Each user and item is associated with real-valued features. For serving, we co-train a learned index with $M$ to replace approximate KNN.

\noindent\textbf{Notations.} $n_i$ denotes the $i$-th node, $X(n_i)$ its features, $N_U(n_i)$ and $N_I(n_i)$ its user and item neighbors. An edge is $(n_i, n_j, w_{n_i,n_j})$ with weight $w_{n_i, n_j}$.

\section{Method}
\label{sec:method}
\subsection{Overview}
\label{sec:overview}
\begin{figure*}[ht!]
    \centering
    \resizebox{\textwidth}{!}{
    \begin{tikzpicture}[
        >=stealth,
        SciBlue/.style={fill=blue!10, draw=blue!70!black},
        SciOrange/.style={fill=orange!10, draw=orange!80!black},
        SciGreen/.style={fill=teal!10, draw=teal!70!black},
        SciRed/.style={fill=red!10, draw=red!70!black},
        SciPurple/.style={fill=violet!10, draw=violet!70!black},
        SciGray/.style={fill=gray!10, draw=gray!60},
        user/.style={circle, SciBlue, thick, minimum size=8mm, inner sep=0pt, font=\small\bfseries},
        item/.style={rectangle, rounded corners=2pt, SciOrange, thick, minimum size=8mm, inner sep=0pt, font=\small\bfseries},
        process/.style={rectangle, rounded corners=4pt, SciGray, thick, minimum height=1.2cm, align=center, font=\small, drop shadow={opacity=0.1, shadow xshift=1.5pt, shadow yshift=-1.5pt}},
        highlight_process/.style={rectangle, rounded corners=4pt, SciGreen, thick, minimum height=1.2cm, align=center, font=\small, drop shadow={opacity=0.1, shadow xshift=1.5pt, shadow yshift=-1.5pt}},
        panel/.style={rectangle, draw=gray!40, fill=gray!3, rounded corners=8pt, inner sep=0pt},
        label/.style={font=\Large\bfseries, text=black, anchor=north west},
        arrow/.style={-{Stealth[scale=1.2]}, thick, gray!80!black},
        darrow/.style={-{Stealth[scale=1.2]}, dashed, thick, gray!80!black}
    ]

    \node[panel, minimum width=8.4cm, minimum height=14.5cm] (panelA) at (0, 0) {};
    \node[label] at ($(panelA.north west) + (0.3, -0.3)$) {A \hspace{4pt} \normalsize Graph Construction \& Subsampling};

    \node[user] (u1) at (-1.5, 5) {$U_1$};
    \node[user] (u2) at (0, 5) {$U_2$};
    \node[user] (u3) at (1.5, 5) {$U_3$};
    \node[item] (i1) at (-0.8, 3.5) {$I_1$};
    \node[item] (i2) at (0.8, 3.5) {$I_2$};
    \draw[thick, gray] (u1) -- (i1);
    \draw[thick, gray] (u2) -- (i1);
    \draw[thick, gray] (u2) -- (i2);
    \draw[thick, gray] (u3) -- (i2);
    \node[font=\footnotesize\itshape, text=gray!80!black] at (0, 5.8) {Raw Interactions ($\sim 100$ Trillion Edges)};

    \node[highlight_process, minimum width=7.4cm] (proc1) at (0, 1.6) {
        \textbf{Edge Subsampling \& Popularity Correction} \\[-2pt]
        \scriptsize $w'_{n_i, n_j} = w_{n_i, n_j} \times \Big(\frac{w_{n_j, n_i}}{\sum_{k}w_{n_j, k}}\Big)^{\alpha}$
    };
    \draw[arrow] (0, 3) -- (proc1);

    \node[user] (u1b) at (-1.5, -1.2) {$U_1$};
    \node[user] (u2b) at (0, -1.2) {$U_2$};
    \node[user] (u3b) at (1.5, -1.2) {$U_3$};
    \node[item] (i1b) at (-0.8, -2.7) {$I_1$};
    \node[item] (i2b) at (0.8, -2.7) {$I_2$};
    \draw[thick, gray] (u1b) -- (i1b) node[midway, left, font=\scriptsize] {U-I};
    \draw[thick, gray] (u2b) -- (i1b);
    \draw[thick, gray] (u2b) -- (i2b);
    \draw[thick, gray] (u3b) -- (i2b);
    \draw[thick, blue!60, dashed] (u1b) to[bend left=25] node[midway, above, font=\scriptsize] {U-U} (u2b);
    \draw[thick, blue!60, dashed] (u2b) to[bend left=25] (u3b);
    \draw[thick, orange!80, dashed] (i1b) to[bend right=25] node[midway, below, font=\scriptsize] {I-I} (i2b);
    \node[font=\footnotesize\itshape, text=gray!80!black] at (0, -0.4) {Extended Heterogeneous Graph};

    \draw[arrow] (proc1) -- (0, -0.1);

    \node[process, minimum width=7.4cm] (proc2) at (0, -4.8) {
        \textbf{Neighborhood Pre-computation} \\
        \scriptsize Personalized PageRank (PPR) + KNN Fallback
    };
    \draw[arrow] (0, -3.2) -- (proc2);

    \node[rectangle, draw=gray!80, fill=white, rounded corners=2pt, dashed, minimum width=7cm, minimum height=0.8cm, font=\small] (outA) at (0, -6.4) {Self-contained Edge-Centric Data};
    \draw[arrow] (proc2) -- (outA);

    \node[panel, minimum width=8.4cm, minimum height=14.5cm] (panelB) at (9, 0) {};
    \node[label] at ($(panelB.north west) + (0.3, -0.3)$) {B \hspace{4pt} \normalsize Heterogeneous Graph Learning};

    \node[process, SciBlue, minimum width=2cm, minimum height=0.8cm] (nu) at (6.5, 4.5) {$N_U(u_i)$};
    \node[user, minimum size=9mm] (target) at (9, 4.5) {$u_i$};
    \node[process, SciOrange, minimum width=2cm, minimum height=0.8cm] (ni) at (11.5, 4.5) {$N_I(u_i)$};
    \node[font=\footnotesize\itshape, text=gray!80!black] at (9, 5.5) {Target Node \& Pre-computed Neighbors};

    \foreach \x in {6.5, 9, 11.5} {
        \node[process, fill=white, minimum width=2cm, minimum height=1cm] at (\x+0.1, 1.9) {};
    }
    \node[process, minimum width=2cm, minimum height=1cm] (encU1) at (6.5, 2) {$f_U$};
    \node[process, minimum width=2cm, minimum height=1cm] (encU2) at (9, 2) {$f_U$};
    \node[process, minimum width=2cm, minimum height=1cm] (encI) at (11.5, 2) {$f_I$};
    \node[font=\scriptsize, align=center] at (6.5, 1.2) {Multi-head \\ User Encoder};
    \node[font=\scriptsize, align=center] at (11.5, 1.2) {Multi-head \\ Item Encoder};

    \draw[arrow] (nu) -- (encU1);
    \draw[arrow] (target) -- (encU2);
    \draw[arrow] (ni) -- (encI);

    \node[highlight_process, SciPurple, minimum width=7cm, minimum height=1.2cm] (agg) at (9, -0.5) {\textbf{Heterogeneous Aggregator} ($AGG_U$)};
    \draw[arrow] (encU1) -- (agg.north -| encU1);
    \draw[arrow] (encU2) -- (agg.north);
    \draw[arrow] (encI) -- (agg.north -| encI);

    \node[process, fill=yellow!20, draw=yellow!60!black, minimum width=4cm, minimum height=1cm] (emb) at (9, -2.5) {\textbf{Final Embedding} $M(u_i)$};
    \draw[arrow] (agg) -- (emb);

    \node[process, minimum width=7cm, minimum height=1cm] (negs) at (9, -4.5) {\textbf{Negative Sampling} \\ \scriptsize In-batch + Out-of-batch + Augmentation};
    \draw[arrow] (emb) -- (negs);

    \node[process, SciRed, minimum width=7.4cm, minimum height=1.2cm] (loss) at (9, -6.3) {
        \textbf{Contrastive Link Prediction Objective} \\[-2pt]
        \scriptsize $L_{i,j} = \lambda L^\text{margin}_{i,j} + (1-\lambda) L^\text{infoNCE}_{i,j}$
    };
    \draw[arrow] (negs) -- (loss);

    \node[panel, minimum width=8.4cm, minimum height=14.5cm] (panelC) at (18, 0) {};
    \node[label] at ($(panelC.north west) + (0.3, -0.3)$) {C \hspace{4pt} \normalsize Index Co-Learning \& Serving};

    \node[process, fill=yellow!20, draw=yellow!60!black, minimum width=3cm, minimum height=1cm] (embC) at (19.5, 4.5) {$h = M(u_i)$};

    \node[highlight_process, SciGray, minimum width=4.5cm, minimum height=1.2cm] (rq1) at (19.5, 2.3) {
        \textbf{Residual Quant. Layer 1} \\[-2pt]
        \scriptsize $k_1 = \arg\min \|h - \mathcal{C}_{1,j}\|^2$
    };
    \node[highlight_process, SciGray, minimum width=4.5cm, minimum height=1.2cm] (rq2) at (19.5, -0.2) {
        \textbf{Residual Quant. Layer 2} \\[-2pt]
        \scriptsize $k_2 = \arg\min \|h_1 - \mathcal{C}_{2,j}\|^2$
    };

    \node[rectangle, draw=teal!80!black, fill=white, thick, minimum width=2.2cm, minimum height=1.8cm, rounded corners=2pt, drop shadow={opacity=0.1}] (cb1_bg) at (15.5, 2.3) {};
    \node[font=\footnotesize\bfseries] at (15.5, 2.9) {Codebook $\mathcal{C}_1$};
    \foreach \x in {-0.7,-0.35,0,0.35,0.7} {
        \foreach \y in {-0.5,-0.15,0.2} {
            \fill[teal!40, rounded corners=0.5pt] (15.5+\x-0.1, 2.3+\y-0.1) rectangle ++(0.2, 0.2);
        }
    }
    \fill[teal!80!black] (15.5-0.1, 2.3-0.25-0.1) rectangle ++(0.2, 0.2); 

    \node[rectangle, draw=teal!80!black, fill=white, thick, minimum width=2.2cm, minimum height=1.8cm, rounded corners=2pt, drop shadow={opacity=0.1}] (cb2_bg) at (15.5, -0.2) {};
    \node[font=\footnotesize\bfseries] at (15.5, 0.4) {Codebook $\mathcal{C}_2$};
    \foreach \x in {-0.7,-0.35,0,0.35,0.7} {
        \foreach \y in {-0.5,-0.15,0.2} {
            \fill[teal!40, rounded corners=0.5pt] (15.5+\x-0.1, -0.2+\y-0.1) rectangle ++(0.2, 0.2);
        }
    }
    \fill[teal!80!black] (15.5+0.25, -0.2+0.1) rectangle ++(0.2, 0.2); 

    \draw[arrow] (embC) -- (rq1);
    \draw[arrow] (rq1) -- node[right, font=\scriptsize] {residual $h_1$} (rq2);
    \draw[arrow] (cb1_bg.east) -- (rq1.west);
    \draw[arrow] (cb2_bg.east) -- (rq2.west);

    \node[process, minimum width=4.5cm, minimum height=1cm] (recon) at (19.5, -2.5) {
        \textbf{Reconstructed Emb.} \\[-2pt]
        \scriptsize $h' = \sum_{i} \mathcal{C}_{i, k_i}$
    };
    \draw[arrow] (rq2) -- (recon);

    \node[process, SciRed, minimum width=7.4cm, minimum height=1.2cm] (reg) at (18, -4.5) {
        \textbf{Code Imbalance Regularization} \\[-2pt]
        \scriptsize Biased Code Selection \& $L^\text{reg} = \hat{p} \cdot p(\mathcal{C})^\text{batch}$
    };

    \draw[arrow] (recon) -- (reg.north -| recon);
    \draw[arrow] (cb2_bg) -- (reg.north -| cb2_bg);
    \node[highlight_process, SciBlue, minimum width=7.4cm, minimum height=1.2cm] (serve) at (18, -6.3) {
        \textbf{Index-based Real-time U2U2I Retrieval} \\[-2pt]
        \scriptsize Recency-filtered Candidates $\mid$ KNN-free (-83\% Cost)
    };
    \draw[arrow] (reg) -- (serve);

    \draw[-{Stealth[scale=1.5]}, line width=2pt, dashed, gray!50] (4.2, 0) -- (4.8, 0);
    \draw[-{Stealth[scale=1.5]}, line width=2pt, dashed, gray!50] (13.2, 0) -- (13.8, 0);

    \draw[darrow, blue!60, rounded corners=4pt] (emb.east) -- (13.5, -2.5) |- (embC.west) node[near end, above, font=\scriptsize, text=black] {\textbf{Co-train Index}};

    \end{tikzpicture}
    } 
    \vspace{-0.5cm}
    \caption{\textbf{The RankGraph-2 system architecture for billion-node graph learning and retrieval.} \textbf{(A)} Graph Construction transforms raw user-item interaction logs into a tractable, heterogeneous graph via popularity bias correction, edge subsampling, and Personalized PageRank (PPR) multi-hop neighbor pre-computation. \textbf{(B)} Model Training leverages multi-head type-aware feature encoders and a heterogeneous aggregator to optimize node representations using a contrastive link prediction objective. \textbf{(C)} Index Co-Learning utilizes residual quantization to jointly train a regularized cluster index, preventing codebook collapse and enabling KNN-free U2U2I retrieval that reduces serving cost by 83\%.}
    \label{fig:RankGraph-2}
\end{figure*}
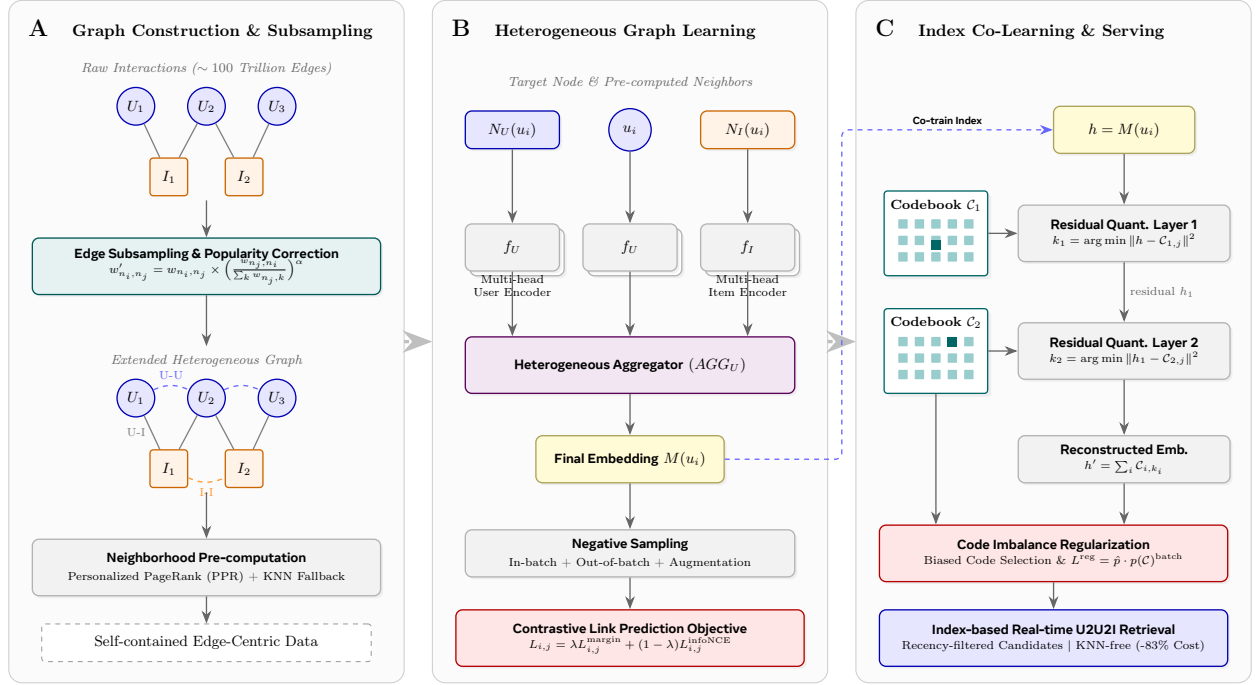

The architecture of RankGraph-2 is shown in Figure~\ref{fig:RankGraph-2}. We focus on the common case of user and item nodes. The following subsections detail the three stages in data-flow order (construction $\rightarrow$ training $\rightarrow$ serving), each shaped by the co-design requirements described in Section~\ref{sec:intro}:
\begin{itemize}[leftmargin=*]
    \item \textbf{Graph construction} (Section~\ref{sec:graph_construction}). Receives requirements from both downstream stages: self-contained edge-centric data with pre-computed neighborhoods (for training), and hour-level rebuild (for serving freshness). Builds a heterogeneous graph with U-U, I-I, and U-I edges from engagement data (Figure~\ref{fig:RankGraph-2}A).
    \item \textbf{Model training} (Section~\ref{sec:training}). Ingests pre-computed data with no online graph access, reusing standard ML infrastructure. Co-learns a cluster index so training directly optimizes for serving quality.
    \item \textbf{Learned index for serving} (Section~\ref{sec:serving}). The co-learned residual-quantization cluster index enables KNN-free U2U2I retrieval with recency filtering, reducing serving cost by 83\%.
\end{itemize}

\subsection{Graph Construction}
\label{sec:graph_construction}
Graph construction receives requirements from both downstream stages: training needs self-contained edge-centric data with pre-computed neighborhoods (no online graph access), and serving needs hour-level graph refresh for item coverage. The graph must also be informative for retrieval while being compact enough to reconstruct within one hour.

\noindent\textbf{Graph schema and edge construction.}
Our graph $G = (V, E)$ has two node types (users $V_U$, items $V_I$) and three edge types (U-I, U-U, I-I), all derived from engagement data (clicks, likes, shares, purchases)---no external social or knowledge graph relations. Each edge carries a scalar weight reflecting engagement strength. U-U and I-I edges are undirected co-engagement edges; U-I edges are directed from user to item. This schema is intentionally simple: we found that a single co-engagement relation per node-type pair, combined with our subsampling strategies, outperforms multiple fine-grained relation types which increase graph size without proportional quality gains. Unlike prior work that uses bipartite U-I graphs~\citep{zheng2018spectral, sharma2024survey, wu2022graph}, we construct all three edge types:
\begin{itemize}[leftmargin=*]
    \item \textit{User-Item (U-I) Edges:} An edge is created between a user node and an item node if the user has engaged with the item within the past $T$ hours. The weight is assigned based on the type of engagement, using predefined values that reflect business value.
    \item \textit{User-User (U-U) Edges:} Let $I(n_i, n_j)$ denote the set of items two user nodes $n_i$ and $n_j$ have both interacted with in the past $T$ hours. An edge is established between the two user nodes $n_i$ and $n_j$ if $|I(n_i, n_j)| >= C_U$, where $C_U$ is a hyperparameter. The weight of the edge is defined as:
    \begin{equation}
        w_{n_i, n_j} = \ln(\sum_{e \in I(n_i, n_j)} w_{n_i, e}*w_{n_j, e})
    \end{equation}
    where we sum the weights induced from each item and then log normalize to keep the weights of frequent users and less frequent users at the same scale.
    \item \textit{Item-Item (I-I) Edges:} The I-I edges are defined in a similar way as U-U edges. Let $U(n_i, n_j)$ denote the set of users two item nodes $n_i$ and $n_j$ have both been interacted by in the past $T$ hours. An edge is established between the two item nodes $n_i$ and $n_j$ if $|U(n_i, n_j)| >= C_I$. The weight of the edge is defined as:
    \begin{equation}
        w_{n_i, n_j} = \ln(\sum_{e \in U(n_i, n_j)} w_{e, n_i}*w_{e, n_j})
    \end{equation}
\end{itemize}

\noindent\textbf{Popularity bias correction.}
Popular items accumulate co-engagement edges that reflect popularity rather than genuine interest, causing over-representation. We adjust I-I edge weights as:
\begin{equation}
    w_{n_i, n_j}' = w_{n_i, n_j} * (\frac{w_{n_j, n_i}}{\sum_{n_k}w_{n_j, n_k}})^{\alpha}
\end{equation}
where $n_k$ iterates over neighbors of $n_j$ and $\alpha=0.3$. Intuitively, if $n_j$ is popular, $\frac{w_{n_j, n_i}}{\sum_{n_k}w_{n_j, n_k}}$ is small, heavily down-weighting the edge. After adjustment, $(n_i, n_j)$ and $(n_j, n_i)$ have different weights; we keep both. Combined with subsampling, this yields far fewer edges from popular items.

\noindent\textbf{Edge subsampling.}
The raw graph has hundreds of trillions of edges. We subsample in two steps: (1) for U-U edges, we retain ${\sim}$0.1B nodes prioritized by business value (refreshed daily); all nodes are retained in U-I edges; (2) for each node, we keep the top $K_{CAP}$ edges by weight. This reduces edges from hundreds of trillions to hundreds of billions. We call the result the extended graph.

Due to edge sampling, nodes partition into two groups: Group~1 nodes have same-type neighbors in the graph; Group~2 nodes do not. Group~1 nodes and their edges form the backbone graph with complete neighbor information. Group~2 nodes appear only in the extended graph.

\noindent\textbf{Pre-computed important neighbors.}
Inspired by PinSage~\citep{ying2018graph}, we use personalized PageRank (PPR)~\citep{yang2024efficient} to identify the $K_{IMP}$ most important user and item neighbors per node. We use a random-walk Monte Carlo approximation ($R$ walks of length $L$, restart probability 0.15), which is embarrassingly parallelizable across billions of nodes. PPR neighbors are \emph{not} added as graph edges; they define a fixed-size pre-computed adjacency list for training, replacing online sampling~\citep{gupta2024graphscale, zheng2022bytegnn}. Edge-type weights are normalized so no type dominates PPR output.

PPR is applied on the backbone graph. For Group~2 nodes (lacking same-type neighbors), we use KNN over Group~1 embeddings from previous training runs (updated daily) to find $K_{IMP}$ same-type neighbors; item neighbors come from top-weight U-I edges. At training time, $K_{IMP}'$ neighbors are randomly sampled from the $K_{IMP}$ pre-computed ones.

\noindent\textbf{Data format.} The output is edge-centric: each record contains the edge $(n_i, n_j, w_{n_i,n_j})$, features and sampled neighbors for both endpoints, partitioned by edge type.

\subsection{Model Training}
\label{sec:training}
Training addresses two co-design responsibilities: it ingests the self-contained edge-centric data from construction (requiring no online graph access), and it co-learns a cluster index to meet serving's KNN-free requirement. The result is a standard supervised link prediction loop that reuses existing ML infrastructure. Our setting is \emph{inductive}: nodes have real-valued features and the model learns shared encoders, generalizing to unseen nodes unlike transductive approaches (e.g., neural matrix factorization).

\noindent\textbf{Model architecture} (Figure~\ref{fig:RankGraph-2}B).
Let $f_t$ ($t \in \{U, I\}$) encode raw features of type $t$ into embeddings. The embedding of node $n_i$ of type $t$ is:
\begin{equation}
\begin{split}
    M(n_i) = AGG_{t}(&f_t(X(n_i)), \\
    &\{f_U(X(e) |e \in N_U(n_i)\}, \\
    &\{f_I(X(e) |e \in N_I(n_i)\}
\end{split}
\end{equation}
where $AGG_{t}$ aggregates user neighbor, item neighbor, and self embeddings. Each node has exactly $K$ user and $K$ item neighbors (fixed during graph construction). Both $f_t$ and $AGG_t$ use multiple heads; multi-head embeddings are used for negative augmentation during training and averaged at inference.

\noindent\textbf{Negative sampling.}
We generate negatives via three strategies: (1)~\emph{in-batch}: negatives from other edges in the same batch; (2)~\emph{out-of-batch}: negatives from a rolling pool across batches to approximate the global distribution; (3)~\emph{negative augmentation}: multi-head embeddings from different heads serve as additional negatives. For each positive edge $(n_i, n_j)$, we sample 100 negatives of the same type as $n_j$.

\noindent\textbf{Loss function.}
We combine margin ranking loss~\citep{schroff2015facenet, dong2018triplet} and infoNCE~\citep{oord2018representation, parulekar2023infonce}. For edge $(n_i, n_j)$, let $s_{i,j}$ be the cosine similarity and $s_{i,k,neg}$ the similarity to negative $n_k$:
\begin{equation}
    L_{i, j}^\text{margin} = \sum_k \max(0, s_{i, k, neg} - s_{i,j}  + \text{margin})
\end{equation}
with margin $= 0.1$. The InfoNCE loss is:
\begin{equation}
    L_{i, j}^\text{infoNCE} = -\text{log}\frac{e^{s_{i,j}/\tau}}{e^{s_{i,j}/\tau} + \sum_k e^{s_{i,k, neg}/\tau}}
\end{equation}
with $\tau = 0.06$. The combined loss per edge is:
\begin{equation}
    L_{i, j} = \lambda L_{i, j}^\text{margin} + (1-\lambda)*L_{i, j}^\text{infoNCE}
\end{equation}
Each batch contains edges from all types. We compute per-type losses $L_{U\text{-}U}$, $L_{U\text{-}I}$, $L_{I\text{-}U}$, $L_{I\text{-}I}$ (U-I is bidirectional). The final loss is:
\begin{equation}
\label{eq:eng}
    \resizebox{\columnwidth}{!}{$L = \beta_1 L_{U-U} + \beta_2*L_{U-I} + \beta_3*L_{I-U} + (1-\beta_1-\beta_2-\beta_3)*L_{I-I}$}
\end{equation}

We adopt the uncertainty weighting method~\citep{kendall2018multi} to learn the hyperparameters $\lambda$, $\beta_1$, $\beta_2$, and $\beta_3$.


\noindent\textbf{Efficiency optimizations.} The graph-infra-free design yields deterministic batch sizes and maximizes Model FLOPs Utilization (MFU)---in contrast to online sampling where multi-hop traversals cause unpredictable memory spikes. We further separate data fetching, feature preprocessing, and negative sampling on CPU, overlapping with on-GPU model training.

\subsection{Learned Index for Efficient Serving}
\label{sec:serving}
Serving is the stage whose cost requirements drive much of the co-design. We serve embeddings for two retrieval scenarios:
\begin{itemize}[leftmargin=*]
    \item \textbf{U2I2I}: Identify a user's engaged items, then find similar items via I2I KNN. Since item embeddings update infrequently, I2I KNN can be pre-computed offline---serving is cheap.
    \item \textbf{U2U2I}: Find similar users, then retrieve their engaged items. We empirically find that constraining the candidate pool to recently active users (e.g., past 15 minutes) yields the best quality. Online KNN over this dynamic pool requires thousands of machines. Pre-computed static KNN mappings are also insufficient, since the active user set changes every few minutes.
\end{itemize}
We resolve this with a cluster-based approach (Section~\ref{sec:intro}): training co-learns a cluster index that assigns each user to a cluster. Each cluster maintains a real-time queue of items from its recently active members. At serving time, U2U2I reduces to reading the latest items from the target user's cluster queue---effectively U2Cluster2I---which is a simple lookup rather than a nearest-neighbor search.

\noindent\textbf{Co-Learned Index.}
To ensure the best cluster quality, we learn the cluster index along with the graph model. We adopt a residual quantization based approach:
\begin{equation}
\begin{split}
    k_i =& \text{arg min}_{j \in [|\mathcal{C}_i|]}||h_{i-1} - \mathcal{C}_{i,j}||^2 \\
    h_i =&  h_{i-1} - \mathcal{C}_{i,k_i}
\end{split}
\end{equation}
where $\mathcal{C}_i$ denote the code book of the $i$-th layer and $\mathcal{C}_{i,j}$ denotes the $j$-th code of the $i$-th layer, $h_i$ is the residual embedding after the $i$-th layer of a node embedding $h$, and $k_i$ is the code of $h$ at layer $i$-th.
The reconstructed embedding for $h$ based on the codebook would be:
\begin{equation}
\label{eq:recon}
    h' = \sum_i C_{i, k_i}
\end{equation}

For a good codebook, the reconstructed embedding should be close to the original embedding, so we define a reconstruction loss for a node embedding:
$L^{\text{recon}} = ||h - h'||^2$.
For each positive edge $(n_i, n_j)$, we obtain the average reconstruction loss over node $n_i$, node $n_j$, and the sampled negative nodes as the reconstruction loss on the edge $L_{i, j}^{\text{recon}}$

Apart from the reconstruction loss, we also use the reconstructed embedding to obtain a loss $L'$ as in Equation~\ref{eq:eng} as the reconstructed embedding should also minimize the graph learning objective.

\noindent\textbf{Cluster-based U2U serving.}
Each user $u_i$ is assigned a hierarchical cluster code $c_i = (k_1, k_2)$ via the residual quantization above. Each cluster maintains a queue of items recently engaged by its active members. At serving time, we simply read the latest items from user $u_i$'s cluster queue---no nearest-neighbor search is needed. Cluster size must be large enough that sufficient recently active users are present after recency filtering, yet small enough to maintain similarity. We tune the codebook sizes ($|\mathcal{C}_1| \times |\mathcal{C}_2|$) to find the best balance. This approach reduces serving cost by 83\% compared to online approximate KNN.

\noindent\textbf{Addressing code imbalance.}
One challenge in learned residual quantization is the imbalance of the learned codebook, i.e. some codes are used much more often than the  others~\citep{kuai2024breaking}. Balancing the codebook and preventing codebook collapse is important in recommendation use cases as the model is always under continuous training.
To address the issue, we propose two techniques:

\textit{(1) Regularization loss}. Ideally the frequency of each code being selected should be equal. To simplify notation, we consider the first layer in the code book.
Let $\hat{p}$ be the empirical probability distribution of the codes in the first layer of the code book in past 1000 batches. We obtain and update $\hat{p}$ by maintaining a queue of code assignments of fixed size 1000.

We then also obtain an estimate in the current data batch of the code selection probability.  For one embedding $h$, its distance to all codes $\mathcal{C}$ is $d(h, \mathcal{C}) = \{d_1, d_2, d_3, \dots \}$. While in the residual quantization step we do hard assignment (i.e. we assign the code with the minimal distance), we perform "soft" assignment here to obtain an approximate probability of the $j$th code being selected as:
\begin{equation}
\label{eq:selection-prob}
    p(h, \mathcal{C})[j] = \frac{e^{\zeta_1/(\zeta_2 + d_j)}}{\sum_k e^{\zeta_1/(\zeta_2 + d_k)}}
\end{equation}
where $\zeta_1=10$ and $\zeta_2=0.01$. Then, the empirical frequency of each code being selected in the current batch can be estimated as:
\begin{equation}
    fre(\mathcal{C})^{\text{batch}} = \{\sum_h p(h, \mathcal{C})[0], \sum_h p(h, \mathcal{C})[1], \dots \}
\end{equation}
and $p(\mathcal{C})^{\text{batch}}$ is obtained by normalizing $fre(\mathcal{C})^{\text{batch}}$.
To push $\hat{p}$ toward uniform, we penalize the current batch for reinforcing already-frequent codes via $L^{\text{reg}} = \hat{p} \cdot p(\mathcal{C})^{\text{batch}}$
averaged over all codebook layers. The final loss combines $L$, $L'$ (contrastive on reconstructed embeddings), $L^{\text{recon}}$, and $L^{\text{reg}}$, balanced via uncertainty weighting~\citep{kendall2018multi}.

\textit{(2) Biased code selection.} Instead of selecting the code that maximizes $p(h_{i-1}, C_i)[j]$, we select:
\begin{equation}
\label{eq:bias-selection}
    k_i = \text{arg max}_{j \in [|\mathcal{C}_i|]} \frac{p(h_{i-1}, C_{i})[j]}{\hat{p}[j]}
\end{equation}
This favors underused codes and penalizes overused ones, preventing codebook collapse under continuous training. The slight dispersion acts as a beneficial exploration mechanism in discovery-oriented surfaces.

\section{Experiments}
\label{sec:experiments}

\subsection{Setup}
\label{sec:setup}

\noindent\textbf{Application context.}
RankGraph-2 is deployed on multiple recommendation surfaces at Meta, and we discuss two major surfaces: Surface~1 and Surface~2, each serving billions of users. The system serves both U2I2I and U2U2I retrieval.

\noindent\textbf{Graph construction.}
The graph is constructed from 24 hours of engagement data and fully reconstructed every 3 hours (${\sim}$1 hour per build). Each graph contains billions of user and item nodes and hundreds of billions of edges across U-U, I-I, and U-I types.

\noindent\textbf{Node features.}
User features include demographics and a sequence of recently engaged items. Item features include content-type and id-based features. All features are encoded into dense embeddings via type-specific encoders.

\noindent\textbf{Model training.}
We train continuously with batch size 32,768, using AdaGrad (lr 0.02) for sparse and AdamW (lr 0.004) for dense parameters. Embedding dimension is 256. The co-learned index uses a two-layer codebook ($5{,}000 \times 50 = 250{,}000$ clusters). We pre-compute $K_{IMP}{=}50$ PPR neighbors per node and sample $K_{IMP}'{=}10$ per training edge.

\noindent\textbf{Embedding refresh.}
After each graph reconstruction, embeddings are regenerated immediately. We aggregate embeddings over the past week (keeping the most recent per node ID). End-to-end latency from engagement data to refreshed embeddings is ${\sim}$2 hours.

\noindent\textbf{Evaluation scope.}
Public benchmarks (e.g., Amazon~\citep{ni2019justifying}, MovieLens~\citep{harper2015movielens}) are orders of magnitude smaller and do not exhibit the scalability challenges that motivate our design. We focus on production-scale evaluation with offline metrics and online A/B tests.

\subsection{Embedding quality evaluation}
\label{sec:embedding_eval}

\subsubsection{Quality of user embedding}
\label{sec:user_eval}
We sample 100K users, retrieve their top-100 KNN neighbors, and measure Recall@$K$ against next-day engagements. Baselines: (1)~\textbf{GAT-DGI}---a Graph Attention Network (GAT)~\citep{velickovic2018graph} with Deep Graph Infomax~\citep{velickovic2019deep} self-supervised pre-training on a user-item bipartite graph, representing a more complex model on a simpler graph; (2)~\textbf{HSTU}~\citep{zhai2024actions}---a trillion-parameter sequential foundation model with embeddings optimized for retrieval via a contrastive objective.

RankGraph-2 outperforms both baselines across all thresholds (Table~\ref{tab:eval_recall}), achieving 3.8$\times$ higher Recall@5 than GAT-DGI. The GAT-DGI comparison is particularly informative: GAT-DGI uses a \emph{more expressive} architecture (GAT + Deep Graph Infomax) but trains on a simpler bipartite graph without heterogeneous co-engagement edges or PPR-based neighbor pre-computation. This gap supports the co-design thesis: the gains come not from model complexity but from lifecycle co-design---since similarity-based retrieval tolerates pre-computed neighborhoods, we invest in construction quality and push serving costs upstream into training. The HSTU comparison further suggests that sequential models, despite massive scale, capture different signals than graph-based structural similarity---consistent with our discussion of complementarity in Section~\ref{sec:discussion}.

\begin{table}[ht!]
\centering
\caption{Eval Recall for RankGraph-2, GAT-DGI, and HSTU on 100,000 sampled users}
\label{tab:eval_recall}
\resizebox{\columnwidth}{!}{%
\begin{tabular}{@{}lllll@{}}
\toprule
Method    & Recall@5 &Recall@10 &Recall@50 &Recall@100  \\ \midrule
GAT-DGI  & 0.038 & 0.066 & 0.180 & 0.251  \\
HSTU & 0.013 & 0.024 & 0.085 & 0.135 \\
RankGraph-2 & 0.143 &0.209 & 0.399 & 0.482 \\
\bottomrule
\end{tabular}
}
\end{table}

\subsubsection{Quality of item embedding}
\label{sec:item_eval}
We evaluate item embeddings on predicting future I-I co-engagement using a strict temporal split: embeddings from day $N$, evaluated against day $N{+}1$ edges. We sample 1,000 edges and measure Recall@$K$ over all pairwise distances. Baselines: (1)~\textbf{PBG}---translational embeddings trained using PyTorch-BigGraph~\citep{lerer2019pytorch} on an item co-engagement graph; (2)~\textbf{HSTU}~\citep{zhai2024actions} as above.

\begin{table}[ht!]
\centering
\caption{Eval Recall for RankGraph-2, HSTU, and PBG on 1000 sampled edges}
\label{tab:eval_recall_item}
\resizebox{\columnwidth}{!}{%
\begin{tabular}{@{}lllll@{}}
\toprule
Method    & Recall@5 &Recall@10 &Recall@50 &Recall@100  \\ \midrule
PBG  & 0.017 & 0.030 & 0.244 & 0.374   \\
HSTU       & 0.132 & 0.177 & 0.349 & 0.434   \\
RankGraph-2  & 0.147 & 0.196 & 0.509 & 0.775 \\
\bottomrule
\end{tabular}
}
\end{table}

RankGraph-2 achieves the highest recall across all thresholds (Table~\ref{tab:eval_recall_item}). At Recall@100, RankGraph-2 achieves 0.775 compared to 0.434 for HSTU (1.8$\times$) and 0.374 for PBG (2.1$\times$). The improvement over PBG is notable because PyTorch-BigGraph is a well-established large-scale graph embedding system; the gap demonstrates that lifecycle co-design---heterogeneous graph construction with PPR-based neighbor selection---substantially outperforms translational embedding approaches that optimize training in isolation.

\subsubsection{Quality of learned index}
We compare original embeddings, reconstructed embeddings (from codebook, Eq.~\ref{eq:recon}), and reconstructed without regularization. Hitrate@$K$ measures whether the positive edge similarity ranks in the top $K$ against negatives.

\begin{table}[ht!]
\centering
\caption{Learned index with vs without the regularization module for code imbalance }
\label{tab:eval_hitrate}
\resizebox{\columnwidth}{!}{%
\begin{tabular}{@{}lllll@{}}
\toprule
Method    & Hitrate@1 & Hitrate@5 &Hitrate@10   \\ \midrule
Original embedding   & 0.871 & 0.981 & 0.997 \\
Recon-embedding   & 0.846 & 0.973 & 0.994 \\
Recon-embedding w/o reg  & 0.717 & 0.877 & 0.931   \\
\bottomrule
\end{tabular}
}
\end{table}

The results (Table~\ref{tab:eval_hitrate}) show that co-learning the index with our regularization techniques maintains high hitrates, closely matching the original embeddings (0.846 vs 0.871 at Hitrate@1). In contrast, removing regularization causes a substantial drop (0.717 at Hitrate@1), demonstrating that codebook collapse is a real risk under continuous training. Codebook utilization confirms this: with regularization, utilization reaches 100\%, whereas without it, utilization drops significantly.

\subsection{Ablation Study}
\label{sec:ablation}
We ablate key design choices using the user embedding evaluation protocol (Recall@$K$, 100K users).

\subsubsection{Effect of heterogeneous edge types}
We compare the full heterogeneous graph against variants with subsets of edge types.

\begin{table}[ht!]
\centering
\caption{Ablation on edge types for user embedding quality}
\label{tab:ablation_edges}
\resizebox{\columnwidth}{!}{%
\begin{tabular}{@{}lllll@{}}
\toprule
Edge Types    & Recall@5 &Recall@10 &Recall@50 &Recall@100  \\ \midrule
U-I only  & 0.052 & 0.081 & 0.210 & 0.285  \\
U-I + I-I   & 0.089 & 0.134 & 0.285 & 0.352 \\
U-I + U-U   & 0.115 & 0.170 & 0.331 & 0.410 \\
U-I + U-U + I-I  & 0.143 & 0.209 & 0.399 & 0.482 \\
\bottomrule
\end{tabular}
}
\end{table}

The bipartite baseline provides a foundation, but I-I edges improve retrieval of long-tail items by encoding direct content similarity, and U-U edges enhance serendipitous discovery outside the user's immediate history. The full topology yields compounding gains that validate the lifecycle investment in heterogeneous graph construction.

\subsubsection{Effect of neighbor selection strategy}

\begin{table}[ht!]
\centering
\caption{Ablation on neighbor selection strategy for user embedding quality}
\label{tab:ablation_neighbors}
\resizebox{\columnwidth}{!}{%
\begin{tabular}{@{}lllll@{}}
\toprule
Neighbor Strategy   & Recall@5 &Recall@10 &Recall@50 &Recall@100  \\ \midrule
Random  & 0.045 & 0.072 & 0.190 & 0.260  \\
Top-weight   & 0.098 & 0.145 & 0.295 & 0.370 \\
PPR neighbors & 0.143 & 0.209 & 0.399 & 0.482 \\
\bottomrule
\end{tabular}
}
\end{table}

``Random neighbors'' randomly samples $K$ neighbors. ``Top-weight neighbors'' selects $K$ neighbors with highest edge weight (single-hop only). Random sampling introduces high variance and noise. Top-weight sampling captures immediate relevance but fails to reach distant, structurally important nodes. PPR captures multi-hop importance, which is critical for the lifecycle: the offline pre-computation cost is amortized across training iterations, and the richer neighborhoods directly improve embedding quality without added model complexity.

\subsubsection{Effect of popularity bias correction}
We evaluate on item embedding quality (Section~\ref{sec:item_eval} protocol).

\begin{table}[ht!]
\centering
\caption{Ablation on popularity bias correction for item embedding quality}
\label{tab:ablation_popularity}
\resizebox{\columnwidth}{!}{%
\begin{tabular}{@{}lllll@{}}
\toprule
Method    & Recall@5 &Recall@10 &Recall@50 &Recall@100  \\ \midrule
w/o correction  & 0.112 & 0.151 & 0.415 & 0.620  \\
w/ correction & 0.147 & 0.196 & 0.509 & 0.775 \\
\bottomrule
\end{tabular}
}
\end{table}

Without correction, embeddings over-index on globally popular items, reducing personalization fidelity for tail users. The correction parameter ($\alpha{=}0.3$) offsets hub-node dominance without artificially fragmenting the graph. Together, the ablation results confirm that each lifecycle component---heterogeneous edges, PPR neighbors, and popularity correction---makes a distinct and complementary contribution to retrieval quality.

\subsection{Online A/B Test}
We conducted 14-day A/B tests on both surfaces at Meta. The treatment group adds RankGraph-2 embeddings as a retrieval source alongside the existing production pipeline, which includes collaborative filtering, content-based, and other embedding-based methods. We report CTR and CVR, the primary business metrics. All improvements are statistically significant ($p < 0.05$).

Tables~\ref{tab:i2i-abtest} and~\ref{tab:u2u-abtest} show the A/B test results. RankGraph-2 achieves up to +0.96\% CTR and +2.75\% CVR for U2I2I, and consistent improvements for U2U2I.

\begin{table}[hbt!]
\caption{A/B test results (14-day, statistically significant)}
\begin{minipage}[t]{0.48\columnwidth}
\centering
\subcaption{U2I2I retrieval}
\label{tab:i2i-abtest}
\resizebox{\linewidth}{!}{%
\begin{tabular}{@{}lll@{}}
\toprule
Surface & CTR & CVR \\ \midrule
Surface 1 & +0.96\% & +2.75\% \\
Surface 2 & +0.33\% & +0.76\% \\
\bottomrule
\end{tabular}%
}
\end{minipage}%
\hfill
\begin{minipage}[t]{0.48\columnwidth}
\centering
\subcaption{U2U2I retrieval}
\label{tab:u2u-abtest}
\resizebox{\linewidth}{!}{%
\begin{tabular}{@{}lll@{}}
\toprule
Surface & CTR & CVR \\ \midrule
Surface 1 & +0.40\% & +1.01\% \\
Surface 2 & +0.20\% & +0.17\% \\
\bottomrule
\end{tabular}%
}
\end{minipage}
\end{table}

\noindent\textbf{Learned index vs.\ online KNN.}
For U2U2I retrieval, we compared the co-learned cluster index against the approximate KNN-based serving system. The learned index achieves comparable CTR and CVR to KNN-based serving while reducing serving infrastructure from thousands of machines to a fraction of that cost---an 83\% reduction. This cost saving is important at Meta's scale, where U2U2I retrieval serves billions of requests daily.

\noindent\textbf{Deployment.}
Based on A/B results, RankGraph-2 is deployed as a graph-based embedding system for content recommendation retrieval at Meta, and has powered \textbf{20+ retrieval launches} across major surfaces. 

\section{Discussion}
\label{sec:discussion}

\noindent\textbf{Where co-design gains come from.}
The 3.8$\times$ recall gain over GAT-DGI and 2.1$\times$ over PBG come not from a more expressive aggregator, but from lifecycle co-design. The key mechanism is that moving neighborhood computation offline does not merely match online sampling quality---it enables \emph{higher} quality by unlocking operations that would be infeasible in an online setting. Specifically, offline pre-computation allows us to: (1) construct richer heterogeneous graphs with U-U, I-I, and U-I co-engagement edges and popularity bias correction, which would be too expensive to maintain in an online graph store; (2) run multi-hop PPR over the full backbone graph to identify structurally important neighbors, whereas online sampling is typically limited to single-hop or shallow random walks due to latency constraints; and (3) invest the engineering effort saved from graph infrastructure into graph construction quality---better edge weighting, subsampling strategies, and neighbor selection. Additionally, pushing serving costs upstream allows training to co-learn a high-quality cluster index, directly optimizing for serving quality rather than relying on post-hoc quantization.

For practitioners, the implication is that the conventional approach, i.e. investing in online graph infrastructure to serve increasingly complex GNN architectures, may be the wrong trade-off for similarity-based retrieval. Instead, investing in offline graph construction quality on standard ML infrastructure can yield both better retrieval quality and lower operational cost. Simpler models also offer advantages in training stability, memory efficiency, and serving latency under continuous training~\citep{ying2018graph, borisyuk2024lignn}.

\noindent\textbf{Complementarity with foundation models.}
Graph models and sequential foundation models (HSTU~\citep{zhai2024actions}, OneRec~\citep{wang2024onerec}, PinFM~\citep{zhou2024pinfm}) are complementary: the former captures global collaborative structure via multi-hop message passing, while the latter excels at temporal intent and short-term dynamics. Neither alone captures both. RankGraph-2 embeddings serve as structural prior features to downstream ranking models (including foundation models), yielding additional gains beyond either paradigm alone. Tighter integration, such as initializing graph embeddings with foundation model representations, is a promising future direction.

\noindent\textbf{Freshness and temporal dynamics.}
RankGraph-2 addresses evolving user dynamics at two timescales. The 3-hour graph reconstruction cycle (each build completing in ${\sim}$1 hour) captures shifted engagement patterns and new items within hours. For U2U2I serving, we empirically find that filtering the candidate pool to recently active users (e.g., past 15 minutes) yields the best quality, providing real-time responsiveness without real-time graph updates.

\noindent\textbf{Limitations.}
Our co-design is optimized for similarity-based retrieval (U2U2I, U2I2I) where structural graph signals dominate; for direct user-to-item retrieval, ranking models with real-time behavioral features may be more effective. Pre-computing neighborhoods offline also precludes tasks requiring query-dependent neighborhood exploration. Additionally, the 3-hour refresh may be insufficient for extremely time-sensitive scenarios.

\section{Conclusion}
\label{sec:conclusion}
We presented RankGraph-2, a framework that co-designs graph construction, training, and serving for billion-node similarity-based retrieval. Requirements cascade across stages: serving needs a co-learned cluster index, training needs self-contained data without online graph infrastructure, and construction must satisfy both while maintaining hour-level freshness. This lifecycle co-design enables a simple architecture to achieve 3.8$\times$ higher recall than GAT + Deep Graph Infomax and 2.1$\times$ higher than PyTorch-BigGraph, while reducing serving cost by 83\%. Deployed at Meta, RankGraph-2 delivers up to +0.96\% CTR and +2.75\% CVR. We believe lifecycle co-design extends to other domains where graph-based systems must operate at scale under serving constraints.

\section*{Acknowledgments}
This work would not be possible without work from the following
contributors: Hang Wang, Jeff Wang, Honghao Wei, Yiyi Pan, Yinglong Xia, Jason Liu, Hanqing Zeng, Gilbert Jiang, Ren Chen, Hunter Song, Yao Zhang, Arthi Suresh, Jiang Li, Hao Lin, Siqi Yan, Yanzun Huang, Hao Wang, Wei Zhao, Liang Zhang, Yuming Liu, Hongming Pu, Harrison Zhao, Ziyi Zhao, Lei Huang, Harry Hai Nguyen, Anand Iyer, Jim Li, Tao Ju, Crystal Jin, Ye Wang, Mingda Li, Cong Zhang, Jun Seok Lee, Zhen Wang, Tian Tong, Prabhjot Singh, Pu Zhang, Keke Zhai, Lillian Zhang, Dang Minh Nguyen, Jiazhou Wang, Emy Sun, Lei Chen, Xiaoxing Zhu, Yuting Zhang, Zhe (Joe) Wang, Daisy Shi He, Min Ni, Bi Xue, Sophia (Xueyao) Liang,  Tao Jia, Yang Cao, Chengye Liu, Pan Chen, Jiacheng Wu, Yucheng Zhang, Shang Huang, Jun Xiao, Max Fan, Lu Zhang, Xinyao Hu, Shilin Ding, Haomin Yu, Ke Pan, Jianhui Wu, Yuanyuan Ding, Haoran Wen, Serena Li, Lizhu Zhang, Jack Zhai, Ke Gong, Rohan Katpelly, Ram Ramanathan, Nipun Mathur, Helen Ma, Deepak Vijaywargi, Divij Pasrija.

\bibliographystyle{assets/plainnat}
\balance{}
\bibliography{sample-base}

\end{document}